\documentclass[journal=jacsat,manuscript=article]{achemso}

\usepackage{amsmath}
\usepackage{amsfonts}
\usepackage{hyperref}
\usepackage{braket}
\usepackage{url}
\usepackage{graphicx}
\usepackage[normalem]{ulem}
\usepackage{color}

\definecolor{my_red}{rgb}{.7,0,0}
\definecolor{my_green}{rgb}{0,.7,0}
\definecolor{my_blue}{rgb}{0,0,.7}
\definecolor{my_orange}{rgb}{1,.5,0}
\definecolor{my_purple}{rgb}{.5,0,.5}

\DeclareUnicodeCharacter{2009}{\ }

\author{Kyle A. Hamer}
\email{khamer3@lsu.edu}
\affiliation[Louisiana State University]
{Department of Physics and Astronomy, Louisiana State University, Baton Rouge, LA 70803}

\author{Aderonke S. Folorunso}
\affiliation[Louisiana State University]
{Department of Chemistry, Louisiana State University, Baton Rouge, LA 70803}

\author{Kenneth Lopata}
\affiliation[Louisiana State University]
{Department of Chemistry, Louisiana State University, Baton Rouge, LA 70803}
\altaffiliation{Center for Computation and Technology, Louisiana State University, Baton Rouge, LA 70803}

\author{Kenneth J. Schafer}
\affiliation[Louisiana State University]
{Department of Physics and Astronomy, Louisiana State University, Baton Rouge, LA 70803}

\author{Mette B. Gaarde}
\email{mgaarde1@lsu.edu}
\affiliation[Louisiana State University]
{Department of Physics and Astronomy, Louisiana State University, Baton Rouge, LA 70803}

\author{Fran\c{c}ois Mauger}
\affiliation[Louisiana State University]
{Department of Physics and Astronomy, Louisiana State University, Baton Rouge, LA 70803}

\title{Tracking Charge Migration with Frequency-Matched Strobo-Spectroscopy}

\begin{document}


\newpage
\begin{abstract}
    We present frequency-matched strobo-spectroscopy (FMSS) of charge migration (CM) in bromobutadiyne, simulated with time-dependent density-functional theory. CM+FMSS is a pump-probe scheme that uses a frequency-matched HHG-driving laser as an independent probe step following the creation of a localized hole on the bromine atom that induces CM dynamics. We show that the delay-dependent harmonic yield tracks the phase of the CM dynamics through its sensitivity to the amount of electron density on the bromine end of the molecule. FMSS takes advantage of the intrinsic attosecond time resolution of the HHG process, in which different harmonics are emitted at different times and thus probe different locations of the electron hole. Finally, we show that the CM-induced modulation of the HHG signal is dominated by the recombination step of the HHG process, with negligible contribution from the ionization step.
\end{abstract}


\newpage
\section{Introduction}

\noindent Understanding the ultrafast motion of electrons within matter is of critical importance for many areas of science and technology. One example is charge migration \cite{calegari2016, worner2017, nisoli2017, li2020} (CM): the coherent motion of a positively-charged electron hole along the backbone of a molecule following a localized ionization event, which can be observed on Angstrom spatial scales and attosecond time scales \cite{goulielmakis2010, calegari2014, kraus2015}. CM is a widely-studied phenomenon due to its potential for understanding and perhaps steering downstream processes such as chemical reactions, photosynthesis, and photovoltaics via charge-directed reactivity \cite{remacle1998, golubev2015, mauger2022}. Since its discovery in the late 1990s \cite{cederbaum1999, breidbach2003}, the study of CM has flourished \cite{lunnemann2008, bredtmann2012, golubev2015, kuleff2016, bruner2017, folorunso2021, mansson2021}, with much of this research performed in recent years \cite{barillot2022, li2022, schlegel2023, yong2022, zhao2022, folorunso2023, yu2023, he2022, hamer2022, matselyukh2022, kobayashi2022, he2023}. 

Despite the many challenges of doing experiments on the attosecond timescale, CM has been measured using several different techniques, including X-ray absorption spectroscopy \cite{goulielmakis2010, matselyukh2022, barillot2022, young2018}, photoelectron spectroscopy \cite{stolow2004, calegari2014, ayuso2017, lara-astiaso2018}, and high-harmonic spectroscopy (HHS) \cite{worner2013, kraus2015, peng2019, he2022}. Due to its inherent sub-femtosecond temporal resolution via the attochirp \cite{mairesse2003, dudovich2006, smirnova2009, azoury2017} of the harmonic radiation, in which different harmonic energies are emitted at different times during the laser cycle, HHS is particularly well-suited to perform time-resolved measurements of ultrafast electron dynamics via a pump-probe scheme. It is useful make a distinction between schemes where the CM is initiated by the same laser field that probes those dynamics, in which the CM dynamics is re-initiated every half-laser cycle~\cite{kraus2015, he2022, he2023}, and schemes where the pump and probe steps are independent \cite{hamer2022}, as discussed here. 

In this paper, we present frequency-matched high-harmonic strobo-spectroscopy of charge migration (CM+FMSS), simulated with time-dependent density-functional theory (TDDFT) \cite{kohn1965, runge1984}. We induce CM dynamics in a bromobutadiyne (BrC$_{4}$H) molecule via the creation of a localized hole on the bromine end of the molecule \cite{folorunso2021, hamer2022}. Following the initiation of the CM dynamics, CM+FMSS uses a delay-dependent, few-cycle HHG-driving laser pulse as an independent probe step to precisely determine the time-dependent location of the electron hole, by tracking the amount of electron density on the bromine atom. The driving laser field is polarized perpendicular to the CM motion, so that it does not drive the electron density. We match the frequency $\omega_{L}$ of the laser to $\omega_{\text{CM}}$ such that the position of the electron hole is the same in each half-cycle of the laser field for any given delay. In our recent work~\cite{hamer2022}, we showed that the CM frequency can be extracted using a different application of HHS, based on creating sidebands in the harmonic spectrum for a broad range of laser frequencies not commensurate with the CM frequency. In the current manuscript, FMSS allows us to go further and perform a time- and space-resolved analysis of the CM dynamics by exploiting the intrinsic time dependence of the HHG process (the attochirp).

\begin{figure*}[!t]
    \centering
    \includegraphics[width=0.65\linewidth]{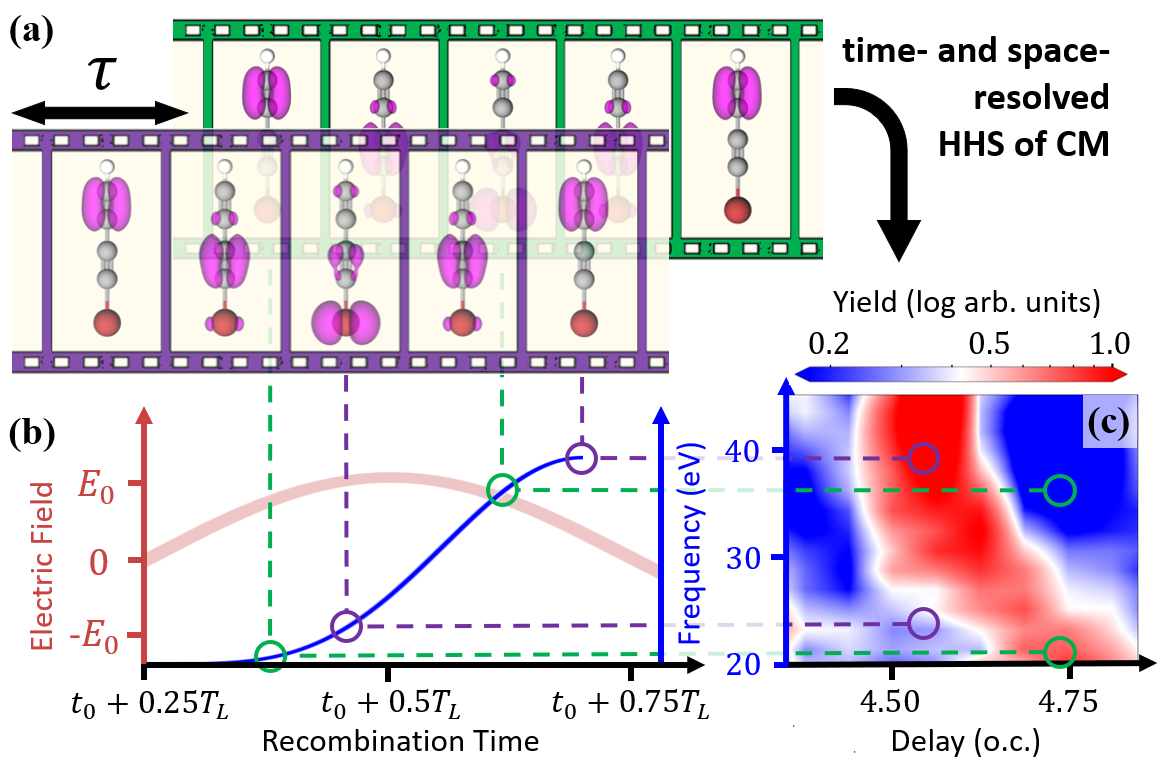}
    \caption{Schematic of our CM+FMSS analysis: (a) snapshots of the time-dependent CM dynamics following the creation of a localized hole on the bromine atom for two different time delays (purple and green frames). (b) These dynamics are probed by a delayed, frequency-matched HHG-driving laser field (red curve). The semiclassical return energy of the rescattered electron wave packet is plotted as a function of the recombination time (blue curve). (c) The resulting orbital-resolved, normalized CM+FMSS spectrum (see text) over half an optical cycle for delays near 4.50 optical cycles (approximately $18\, \text{fs}$ after the initiation of the CM). There is a clear variation in the delay-dependent harmonic spectrum due to the CM dynamics.}
    \label{fig:1}
\end{figure*}

In Figure~{\ref{fig:1}}, we show a schematic which describes the FMSS concept: panel (a) depicts the time-dependent CM dynamics at two different time delays after initiation (purple and green frames, respectively). This dynamics is probed after some time $\tau$ by a HHG-driving laser field, shown in (b), with $\omega_{L} = \omega_{\text{CM}}/2$. Also in panel (b) we show the semiclassical \cite{corkum1993, schafer1993, lewenstein1994, chang2011} time-dependent return energies during one-half cycle of the HHG-driving laser field, which maps to the sub-cycle time-dependent emission frequencies in the harmonic spectrum \cite{varju2005, niikura2005, smirnova2009}. For different delays, a given harmonic energy is emitted at a different time during the CM period, and this strongly affects the resulting HHG yield as shown in panel (c). For example, at a delay of 4.55 optical cycles (o.c.), the low-order (high-order) harmonics are emitted when the hole is located on the bromine atom (terminal $\text{C}\equiv\text{C}$ bond), which gives rise to low (high) HHG yield -- see the purple dashed lines in Fig.~{\ref{fig:1}}. A central finding of this work is that the delay-dependent harmonic yield tracks the time-dependent electron density on the bromine atom, from which we determine the phase of the CM motion. 

\section{Methods}

\noindent In our CM+FMSS simulations, we start by creating a one-electron valence hole localized on the halogen end of a bromobutadiyne ($\text{BrC}_{4}\text{H}$) molecule. We use constrained density functional theory to create an outer-valence hole at $t=0$ that induces particle-like CM along the backbone of the molecule, with a fundamental frequency of $\omega_{\text{CM}} = 1.85\  \text{eV}$, similar to Refs.~\cite{folorunso2021, hamer2022}. This initial hole emulates a halogen-localized ionization \cite{sandor2019, barillot2022}, which leads to CM in the valence shell of the molecule, as we have demonstrated previously~\cite{bruner2017, folorunso2021, mauger2022, hamer2022}. Our preliminary modeling of how to initiate this type of localized CM using a realistic pump pulse is encouraging: we find that both attosecond XUV pulses and few-femtosecond intense infrared pulses initiate similar modes of CM as those discussed in this paper, although we have yet to implement a full calculation including both a realistic pump along with the FMSS-driving pulse.

\begin{figure*}[!t]
    \centering
    \includegraphics[width=0.65\linewidth]{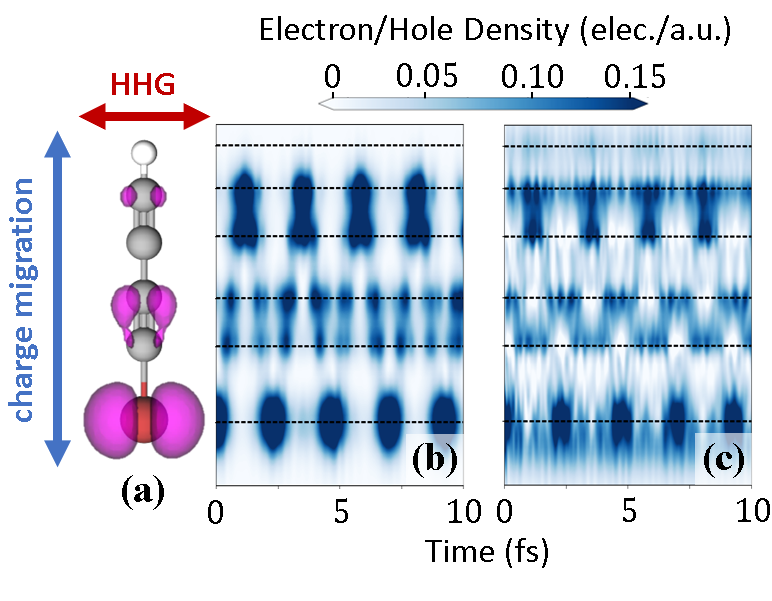}
    \caption{(a) Schematic of the FMSS configuration to probe CM in BrC$_{4}$H. The HHG driving field is polarized perpendicular to the molecular backbone along which the periodic CM occurs. We also show the isosurface of the initial CM orbital density.
    (b,c) Time evolution of the (b) CM orbital density and (c) electron hole density, integrated over the directions transverse to the molecular backbone, as a function of position.}
    \label{fig:2}
\end{figure*}

The CM dynamics is illustrated in Fig.~{\ref{fig:2}}: in panel (a) we show the isosurface of the electron density contribution from the unpaired Kohn-Sham channel from which we remove one electron, here called the \emph{CM orbital} $\psi_{\text{CM}}$. In panels (b) and (c), we show two different representations of the resulting CM dynamics. Fig.~{\ref{fig:2}}(b) shows the time evolution of the CM orbital density, denoted $|\psi_{\text{CM}}(z, t)|^{2}$, integrated over the directions transverse to the molecular backbone. Here, we see a clear oscillation of the electron density in the CM orbital which begins on the bromine atom, travels through the central carbon bond to the terminal bond, and then back again, with a period of $2.24\ \text{fs}$. In Fig.~{\ref{fig:2}}(c), we show the corresponding hole density, defined as the time-dependent density difference between the neutral and the cation densities, $\rho_{h}(z, t) = \rho_{\circ}(z) - \rho_{+}(z, t)$ \cite{cederbaum1999, breidbach2003, folorunso2021}, again integrated over the directions transverse to the molecular backbone. This hole density exhibits a similar pattern to that of the electron density in the CM orbital with additional high-frequency oscillations localized around each of the atomic centers \cite{mauger2022}. In all TDDFT calculations shown in this paper, nuclear dynamics have been omitted.

We next induce HHG in the BrC$_{4}$H cation undergoing CM, using a laser pulse with a polarization direction perpendicular to the molecular backbone so that we do not drive the electron density along the molecular backbone. This laser pulse has a frequency $\omega_{L} = \omega_{\text{CM}}/2$ (corresponding to a laser wavelength $\lambda_{L} = 1344\ \text{nm}$). The frequency-matching condition is chosen such that the electron hole is at the same position along the molecular backbone at every half-cycle of the laser field. By using different sub-cycle delays between the initiation of the CM and the laser field, we therefore sample different positions of the electron hole along the molecular backbone. For our TDDFT simulations, we use $\sin^{2}$ laser pulses centered around a delay $\tau$ relative to the initiation of the CM, and that last for 5 o.c. in total ($\approx 1.5$ o.c. FWHM). We then scan the sub-cycle-resolved delay over two full laser cycles, advancing the delay $\tau$ in increments of $1/16$ optical cycles. In all simulations we use a peak intensity of $45\ \text{TW}/\text{cm}^{2}$, leading to a cutoff energy of around $40\ \text{eV}$.

We use grid-based TDDFT with a local-density-approximation exchange-correlation functional \cite{perdew1981, yabana1996, marques2012} and average-density self-interaction correction \cite{legrand2002, ciofini2003, tsuneda2014} within the OCTOPUS software package \cite{marques2003, castro2006, andrade2015, tancogne-dejean2020} to describe both the CM and HHG processes. We use a simulation box with dimensions of $90 \times 40 \times 90\  \text{a.u.}$ (with the shorter box length transverse to both the laser field and molecular axes), and a complex absorbing potential that extends $15\  \text{a.u.}$ from each edge of the box. We choose the box dimensions such that we select the short-trajectory contribution to the HHG spectrum that is usually observed in HHG measurements \cite{hamer2021}, by absorbing the long-trajectory contribution. We use a grid spacing of $0.3\  \text{a.u.}$ in all directions. 

In order to compute harmonic spectra, we first define the orbital-resolved dipole moment \cite{hamer2021} $d_{j}(t)$ corresponding to the $j^{\text{th}}$ Kohn-Sham orbital $\psi_{j}(\vec{r}, t)$:
\begin{equation}\label{eq:ordm}
    \vec{d}_{j}[\tau](t) \cdot \hat{x} = \int x\, \mathrm{d}x \iint \mathrm{d}y\, \mathrm{d}z\, |\psi_{j}[\tau](\vec{r}, t)|^{2}
\end{equation}
Here, we focus on the dipole signal parallel to the driving laser field (in the $x$-direction). We have checked that our results are nearly identical when including the dipole signal in the directions perpendicular to the laser field. The oscillating charge density along the molecular axis induces a significant dipole contribution along the axis of the molecular backbone, as was also observed in Kuleff \emph{et al}~\cite{kuleff2011}. However, above 20 eV, the total emission spectrum is dominated by the driven (harmonic) response, with the CM-only emission rapidly decreasing with respect to the emission frequency. In the remainder of this paper, we focus on either the combined dipole signal from the three $\pi$ orbitals in which the CM takes place (and where the vast majority of the harmonic signal resides), or the dipole signal from only the CM orbital defined above. We window the thus-computed dipole moment in the time domain using a $\cos^{2}$ function which has the same width as the laser pulse, such that the time-dependent signal smoothly goes to zero on both ends. Then, we apply a Fourier transform, and square to obtain the delay-dependent HHG yield:
\begin{equation}\label{eq:fmss_spec}
    S[\tau](\omega) = \bigg{|} \mathcal{F} \left[W(t, \tau) \cdot \left(\vec{d}[\tau](t) \cdot \hat{x}\right)\right] \bigg{|}^{2}
\end{equation}
Lastly, in order to more clearly investigate the CM-induced delay-dependent modulation of the harmonic signal, we first smooth the spectrum using a moving average to remove the individual harmonic peaks and then normalize $S[\tau](\omega)$ by the delay-averaged harmonic signal; this final step removes the general shape (perturbative region, plateau, and cutoff region) of the harmonic spectrum and focuses on the delay dependence.

\section{Results and Discussion}

\begin{figure}[!t]
    \centering
    \includegraphics[width=0.65\linewidth]{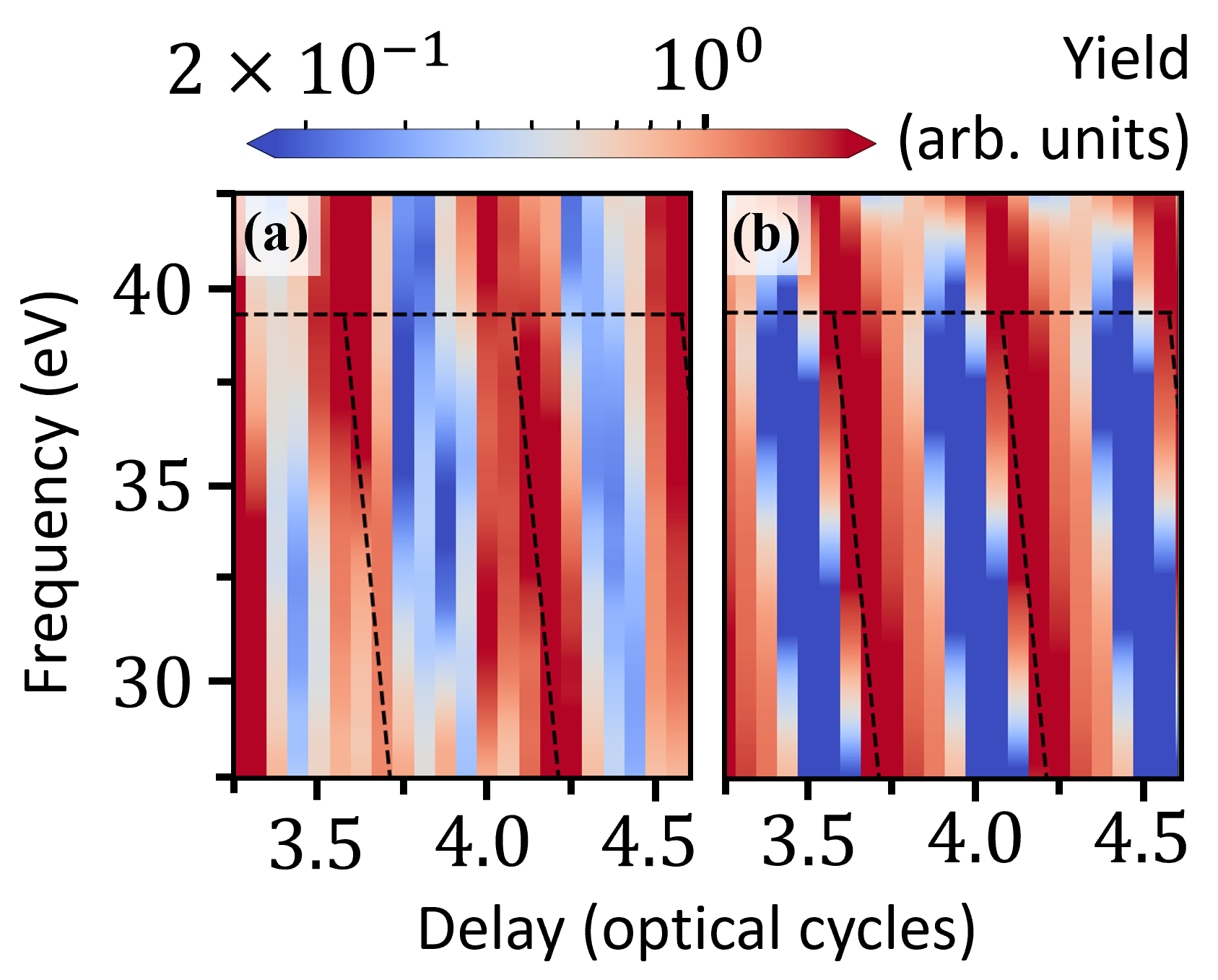}
    \caption{(a) Normalized, CM-orbital-resolved FMSS spectrum for $\lambda_{L} = 1344\  \text{nm}$ ($\omega_{L} = \omega_{\text{CM}}/2$), $I_{\circ} = 45\  \text{TW}/\text{cm}^{2}$. (b) Delay-dependent harmonic spectrum taken from model calculations (see text). Black dashed lines are taken from ridge detection of the peaks in panel (b).}
    \label{fig:3}
\end{figure}

\noindent In Figure~\ref{fig:3}(a), we show the normalized CM+FMSS spectrum calculated from the CM-orbital-resolved dipole moment described by Eq.~{\ref{eq:ordm}}, also shown previously in Fig.~{\ref{fig:1}}(c), around 4 optical cycles (approximately $18\, \text{fs}$) after the initiation of the CM. Clearly, there is a pronounced half-laser-cycle-periodic, delay- and harmonic-frequency-dependent variation in the harmonic signal which is not present in the absence of the CM dynamics (in the neutral molecule). This variation is such that the yield is roughly five times more intense when the hole is \emph{not} on the bromine atom. Below the cutoff energy $E_{c} = 40\ \text{eV}$, this spectral maximum trends towards earlier delays as the harmonic frequency increases. As we discuss below, the slope of this tilt matches the negative of the attochirp of the harmonic radiation.

To further investigate our TDDFT results, we construct a CM-modulated model HHG dipole moment based on the strong-field approximation (SFA) \cite{lewenstein1994}. In the absence of CM, the idealized harmonic response from a gas-phase target irradiated by a monochromatic laser field with a frequency $\omega_{L}$, a cutoff frequency $\omega_{c}$, and an envelope $F(t)$ is given by
\begin{equation}\label{eq:dip_noCM}
    d(t) = F(t) \cdot \sum_{n=1,3,5\ldots}^{\infty} A_{n} \sin{\left(n\omega_{L}t + \phi_{n}\right)}
\end{equation}
where the amplitude and phase of the $n^{\text{th}}$ harmonic are respectively defined by
\begin{align}
    A_{n} &= \begin{cases} 1, & n \omega_{L} \leq \omega_{c} \\ e^{-(n\omega_{L} - \omega_{c})}, & \text{otherwise} \end{cases} \label{eq:model_amp}\\
    \phi_{n} &= -\frac{3 \omega_{L}}{10 U_{p}}n^{2} - n \omega_{L} \tau \label{eq:model_phase}
\end{align}
Here, $U_{p}$ is the ponderomotive energy and $\tau$ is the time delay defined above. The harmonic phases $\phi_{n}$ are defined such that we replicate a linear approximation of the semiclassical short-trajectory attochirp \cite{chang2011}. To model the effect of the CM dynamics, we modulate the signal in Eq.~{\ref{eq:dip_noCM}} via:
\begin{equation}\label{eq:dip_CM}
    \tilde{d}(t) = d(t) \cdot \left[1 + \sum_{m} B_{m} \sin{(m \omega_{\text{CM}} t + \phi_{m})}\right]
\end{equation}
where the parameters $\{m, B_{m}, \phi_{m}\}$ describe the individual Fourier components of the field-free CM dynamics.  Consistent with our previous results \cite{hamer2022}, we include two main Fourier components in the CM dynamics: one at $\omega = 1 \cdot \omega_{\text{CM}}$, and a second at $2 \cdot \omega_{\text{CM}}$ which is roughly four times less intense than the first, and with an extra $\pi/4$ phase shift (see again Fig.~{\ref{fig:2}}). 

We plot the delay-dependent harmonic spectrum calculated from the model dipole signal of Eq.~{\ref{eq:dip_CM}} in Figure~{\ref{fig:3}}(b). Like in panel (a), we see a half-cycle-periodic modulation tilting to the left as the harmonic frequency increases. The modulations seen in both panels are consistent with one another, as evidenced by the black dashed lines in both plots, taken from a ridge detection of the peaks in the model spectrum in (b). Removing the attochirp from our model calculations (first term in Eq.~{\ref{eq:model_phase}}) eliminates the slope of the variation shown in Fig.~{\ref{fig:3}}(b), since in the absence of the attochirp all electron trajectories return at the same time regardless of harmonic frequency. The delay dependence of the variation in the harmonic signal in Fig.~{\ref{fig:3}} is therefore sensitive to the attochirp of the harmonic radiation, as illustrated in Fig.~{\ref{fig:1}}. The rescattered electron wave packet images different molecular landscapes depending on when it rescatters \cite{varju2005, niikura2005, smirnova2009, dudovich2006, azoury2017}, leading to a variation in the HHG light emission. High-frequency light (near the cutoff energy) is emitted later, meaning that an earlier delay is required to image any given position of the hole along the molecular backbone. Note that harmonic generation from any neutral molecules not undergoing CM would not have any delay dependence, and so would be canceled out by the normalization process. We also note that the time resolution built into FMSS via the attochirp means that there will be a delay and frequency dependence to the harmonic yield even if $\omega_{CM}$ does not match $\omega_{L}/2$ exactly, {\it i.e.}, as long as $1/|\omega_{CM}-\omega_{L}/2|$ is small compared to the time (delay) duration over which the CM is sampled.

From the purple and green dashed lines in Fig.~{\ref{fig:1}}, we see that the HHG yield increases when the hole is located in the terminal bond (\emph{i.e.} when the electron density is on the bromine atom), and vice versa. This conclusion suggests that the scattering cross-section of the bromine atom is larger than the rest of the carbon chain, meaning that an increase in the overall density on the bromine atom (when the hole is \emph{not} on the halogen) results in a relative increase in the harmonic yield. This is a crucial result: because there is a spatially-resolvable feature in the harmonic spectrum -- here, a decrease in the harmonic yield when the hole is located on the halogen atom -- we are able to perform a time- and space-resolved analysis of the CM dynamics using FMSS.

Though we are simulating and measuring particle-like CM dynamics \cite{folorunso2021, hamer2022} in this work, we expect that FMSS can be used to characterize a variety of ultrafast electron dynamics. The only requirement is that there is one or more features of the harmonic yield that can be traced back to specific parts of the molecule. As an example, in the usual way that CM is described, as a back-and-forth motion between two sites (\emph{e.g.} bromoacetylene), a measure of the amount of electron density on one of the sites fully describes the CM motion since any hole density not on the probed site must be on the other site. 

\begin{figure}[!t]
    \centering
    \includegraphics[width=0.65\linewidth]{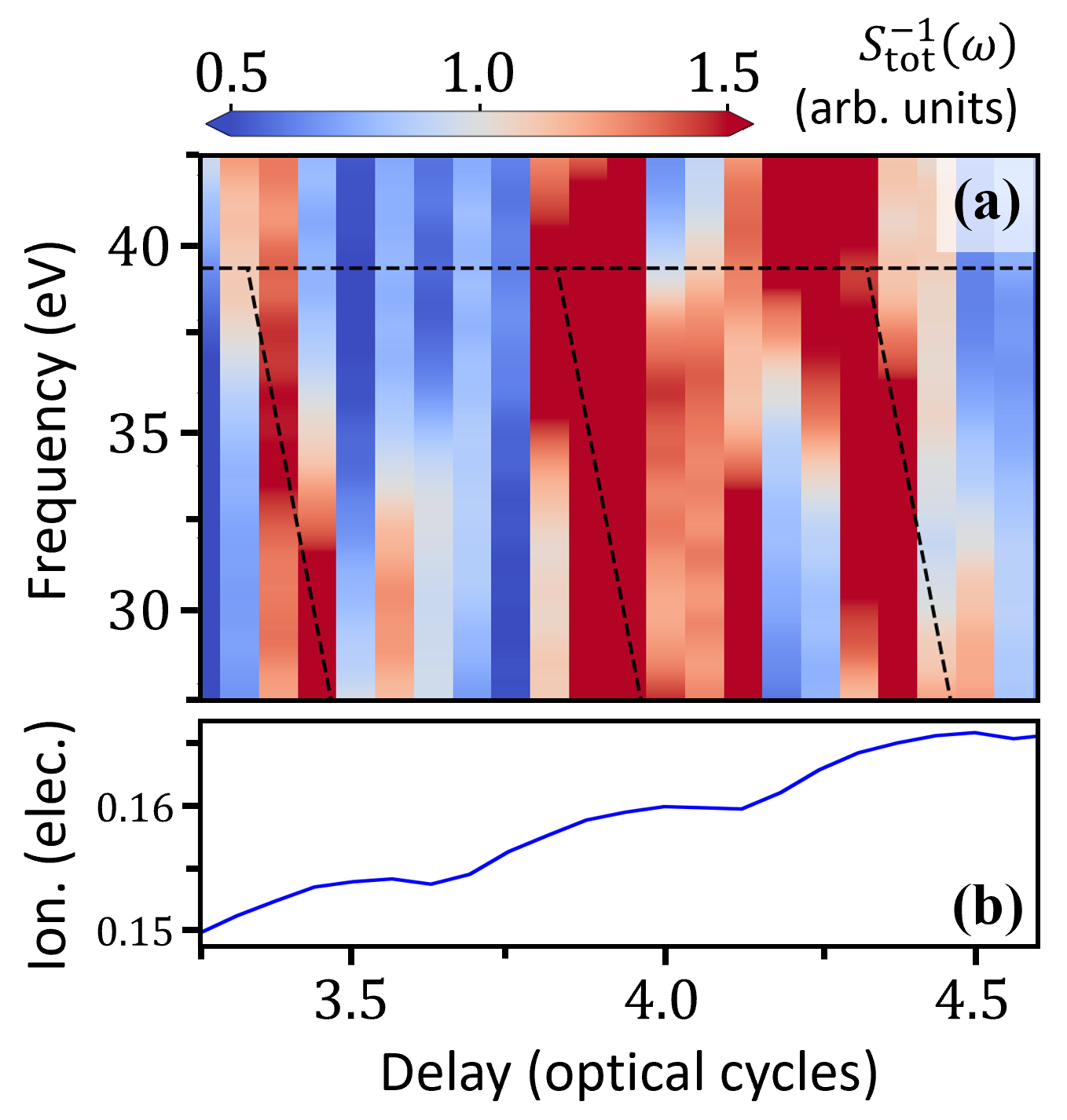}
    \caption{(a) Normalized CM+FMSS spectrum, again for $\lambda_{L} = 1344\  \text{nm}$ and $I_{\circ} = 45\  \text{TW}/\text{cm}^{2}$, calculated from the dipole signal from the three $\pi_{xz}$ orbitals. Black dashed lines again taken from model calculations. (b) Amount of ionized charge, three optical cycles after the center of the laser pulse, as a function of delay.}
    \label{fig:4}
\end{figure}

While the CM orbital used in Fig.~{\ref{fig:3}} gives us the clearest picture of the CM dynamics (see again Fig.~{\ref{fig:2}}(b)), it does not correspond to a physical observable. Consider the electronic structure of $\text{BrC}_{4}\text{H}$: in addition to some lower-lying $\sigma$-type orbitals that do not contribute to the CM or the HHG, there are six $\pi$-type orbitals that span the length of the molecular backbone. Three of these $\pi$ orbitals lie in the $xz$-plane (where the molecular backbone is along the $z$-axis, and the laser is along the $x$-axis), while the other three lie in the $yz$-plane. By pulling one of the two electrons from one of the $\pi$ orbitals in the $xz$-plane (the CM orbital), we induce particle-like CM in $\text{BrC}_{4}\text{H}$; however, there are an additional 4 electrons in the $\pi_{xz}$ system that strongly contribute to both the CM and the HHG. Therefore, we look at the combined dipole signal from the three $\pi_{xz}$ orbitals. We have checked that these results are consistent with using the total dipole acceleration rather than the $\pi_{xz}$-orbitals-resolved dipole moment. 

Thus far, we have been looking at the relative increase in the delay-dependent HHG yield that occurs when the hole is $\emph{not}$ on the bromine atom. This method works well for the CM-orbital-resolved FMSS spectrum of Fig.~{\ref{fig:3}}(a); switching to the $\pi_{xz}$-orbitals-resolved FMSS spectrum, however, we instead look for an \emph{absence} of harmonic yield corresponding to the hole being on the bromine atom. Thus, in Fig.~{\ref{fig:4}}(a), we plot the inverse of the $\pi_{xz}$-orbitals-resolved harmonic yield, $1/S_{\pi}(\omega)$. Again, we see a delay- and harmonic-frequency-dependent variation in the (inverse) harmonic yield due to the CM dynamics. The black dashed lines, again taken from our model calculations in Fig.~{\ref{fig:3}}(b), have been shifted by $0.25\omega_{L}$ since we are looking for an absence, rather than the presence, of harmonic signal.

\begin{figure}[!t]
    \centering
    \includegraphics[width=0.65\linewidth]{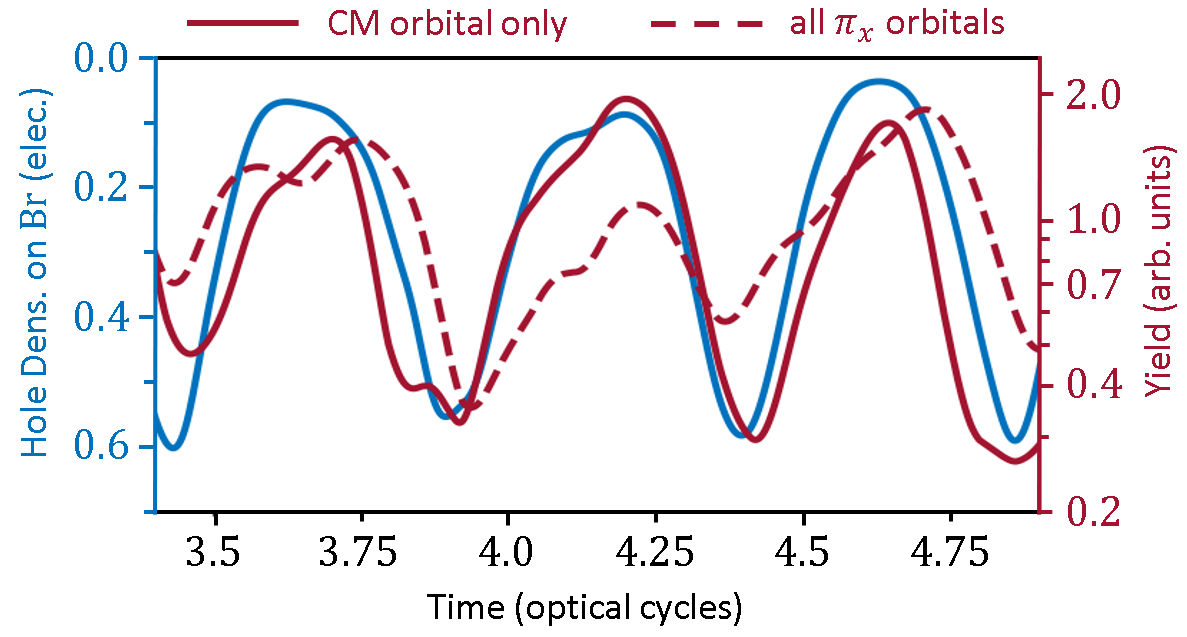}
    \caption{Comparison between time-dependent, field-free hole density on the bromine atom (blue) and recombination-time-dependent HHG yield, with the attochirp removed and integrated over harmonic frequencies above $20\ \text{eV}$ (red). The solid red line is taken from the CM-orbital-resolved spectra in Fig.~{\ref{fig:3}(a)}, and the dashed line corresponds to the $\pi_{xz}$-orbitals-resolved data in Fig.~{\ref{fig:4}(a)}.}
    \label{fig:5}
\end{figure}

We have shown that the HHG yield tracks the hole density on the bromine atom. To further illustrate this, we algebraically remove the effect of the attochirp in the CM+FMSS spectra of Figs.~{\ref{fig:3}}(a) and \ref{fig:4}(a) in order to obtain an absolute-time-dependent measure of how much hole density is on the bromine end of the molecule. This analysis is performed in Figure~{\ref{fig:5}}. The blue curve depicts the amount of hole density centered around the bromine atom, taken from the field-free CM dynamics depicted in Fig.~{\ref{fig:2}}(b). We compare this hole density to the recombination-time-dependent harmonic yield, integrated over harmonic frequencies above 20 eV, for the CM-orbital-resolved data in Fig.~{\ref{fig:3}}(a) (solid red curve) and the $\pi_{xz}$-orbitals-resolved data in Fig.~{\ref{fig:4}}(a) (dashed red curve). From the semiclassical model of HHG \cite{corkum1993, schafer1993, lewenstein1994, chang2011}, we know exactly when each harmonic is emitted as a function of absolute time (for every delay $\tau$). From our TDDFT simulations, we also know the exact location of the electron hole as a function of absolute time. Thus, we can unambiguously map the variation in the harmonic signal to the amount of electron density on the halogen atom. In Fig.~{\ref{fig:5}}, a value near the top of the figure means the hole density is \emph{not} localized on the bromine atom (is localized on the terminal bond), and therefore results in a larger HHG yield. Despite the different methods used to obtain the red and blue curves in Fig.~{\ref{fig:5}}, they match each other very well. Note that the higher-frequency oscillations in the red curves (particularly, the dashed red curve) can be explained by the additional atomic-center-localized oscillations in the hole density in Fig.~{\ref{fig:2}}(c).

We finally ask the question of whether the CM-induced modulation of the harmonic yield is due to the ionization step or the rescattering step. To do so, in Figure~{\ref{fig:4}}(b) we plot the amount of charge ionized from the simulation box, one-half laser cycle after the end of the laser pulse, as a function of the delay $\tau$. There is a small amount of leakage -- ionized charge leaving the simulation box even in the absence of the laser field, approximately 2\% of an electron per laser cycle -- as evidenced by the overall slope in Fig.~{\ref{fig:4}}(b), which can be attributed to the absorbing boundaries in the direction perpendicular to both the CM and the laser. We have checked that the leakage does not effect the results shown here. On top of this overall linear slope, we see a clear a half-laser-cycle-periodic modulation in the ionization signal due to the CM dynamics. Different relative phases between the CM and the peaks of the laser field cause different amounts of charge to be ionized as a function of $\tau$. However, after correcting for the leakage, the amplitude of the oscillation in the ionization signal is quite small (roughly 1\%) compared to the variation in the harmonic signal, for which the yield is roughly three times larger when the hole is not on the bromine atom (as opposed to five times larger, for the CM-orbital-resolved case). Thus, we conclude that the modulation of the HHG signal from the CM dynamics occurs mainly as a result of the recombination step, not the ionization step.

\section{Conclusions}

\noindent In summary, we have shown that frequency-matched high-harmonic strobo-spectroscopy of charge migration (CM+FMSS) in $\text{BrC}_{4}\text{H}$ causes a coherent modulation of the HHG signal that precisely tracks the amount of electron density on the bromine atom, which tells us the phase of the CM motion. By exploiting a site-specific feature of the HHG spectrum, we achieve a time- and space-resolved analysis of the CM by performing a sub-cycle-resolved delay scan. FMSS takes advantage of the intrinsic attosecond time resolution of the HHG process (the attochirp), in which different harmonics are emitted at different times and thus probe different locations of the electron hole. These claims are supported by a similar result from an SFA-inspired model calculation. We can also make a direct comparison between the recombination-time-dependent, harmonic-frequency-integrated HHG yield and the hole density on the halogen.

It is interesting to consider how the FMSS envisioned in this paper would fare when considering more realistic experimental conditions, in particular the two approximations we are making concerning (i) the (perfect) perpendicular alignment of the molecule relative to the laser polarization, and (ii) the absence of nuclear motion. 
For (i), we expect the biggest issue to be that a laser field component that is parallel to the molecular backbone will drive CM that is not necessarily in phase with the field-free CM, and which will therefore likely give rise to a different delay dependence. For bromobutadiyne interacting with the few-cycle laser pulse we have used here, we find that the harmonic response to a parallel-polarized laser pulse does indeed exhibit a different delay dependence but that it is also substantially weaker than that of the perpendicular-polarized pulse and thus does not contribute much in the total delay dependence. For the longer driving pulses used in Ref.\cite{hamer2022}, we found that the sideband-based HHS proposed in that paper was valid for a full-width half-maximum angular distribution of 40$^{\circ}$. Given the weaker response for the shorter pulse duration used here, we expect that FMSS will also tolerate at least 40$^{\circ}$ of angular distribution.

For (ii) we can estimate the effect of including nuclear motion in several different ways. First, we have performed preliminary calculations of CM in bromobutadiyne when including Ehrenfest dynamics and find  that the molecule is quite rigid. A complete characterization of the effect of nuclear motion,  scanning over initial geometries as well as the sub-cycle-resolved delay, is currently computationally intractable when also calculating the HHG spectrum. However, it is also useful to think about the time scale for the nuclear dynamics compared to that of the short pulse and few cycles of probing that we discuss here. In particular, by incorporating decoherence into the model calculations described above we find that FMSS remains applicable within the typical time scale for nuclear dynamics and decoherence.

Beyond the $\text{BrC}_{4}\text{H}$ molecule used here, we note that similar particle-like CM dynamics have been predicted in other classes of molecules \cite{folorunso2021, folorunso2023}. Thus, given the generalizable nature of our approach, we expect that CM+FMSS analyses can be applied broadly to other classes of molecules, such as functionalized benzenes or even bio-molecules and beyond. Given the intense, current interest in probing and understanding charge migration, with a range of experiments underway at large-scale X-ray facilities\cite{li2022, grell2023}, approaches based on HHS, such as FMSS, could be appealing due to the much wider availability of table-top based HHG sources.

\section{Supporting Information}

The TDDFT simulation data and Python scripts we use to produce the figures are available at [authors will provide link to public repository for production].

\section{Acknowledgments}

We thank L. F. DiMauro and R. R. Jones for helpful discussions on this work.
This work was supported by the U.S.\ Department of Energy, Office of Science, Office of Basic Energy Sciences, under Award No.~DE-SC0012462.
Portions of this research were conducted with high performance computational resources provided by Louisiana State University (\url{http://www.hpc.lsu.edu}) and the Louisiana Optical Network Infrastructure (\url{http://www.loni.org}).

\singlespacing

\providecommand{\latin}[1]{#1}
\makeatletter
\providecommand{\doi}
  {\begingroup\let\do\@makeother\dospecials
  \catcode`\{=1 \catcode`\}=2 \doi@aux}
\providecommand{\doi@aux}[1]{\endgroup\texttt{#1}}
\makeatother
\providecommand*\mcitethebibliography{\thebibliography}
\csname @ifundefined\endcsname{endmcitethebibliography}
  {\let\endmcitethebibliography\endthebibliography}{}

\newpage
\section{TOC Graphic}
\hfill

\begin{figure}[!h]
    \centering
    \includegraphics[width=3.25in]{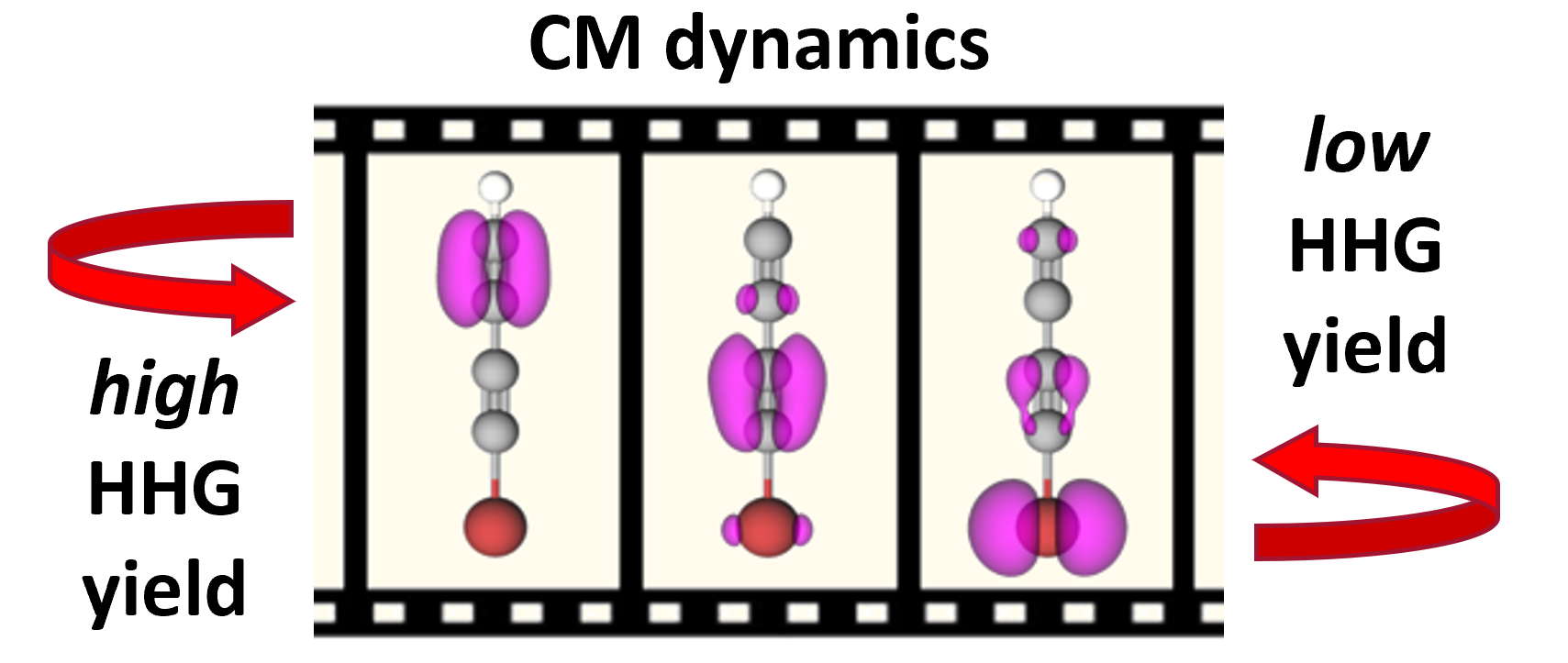}
    \label{fig:toc}
\end{figure}

\end{document}